\documentclass[a4paper]{article}

\usepackage[T1]{fontenc}
\usepackage[dvips]{graphicx}
\usepackage{graphicx,color}
\usepackage{fullpage}
\title{\bf S-process elements in extremely metal-poor stars}
\date{}

\begin{document}
\maketitle
\mbox{}
\begin{center}
{\bf Monique Alves Cruz}
\end{center}
\begin{center}
Max-Plack Institute for Astrophysics, \\Karl-Schwarzschild-Str.1,
85748 Garching, Germany\\
mcruz@mpa-garching.mpg.de
\end{center}
\begin{center}
{\bf Silvia Rossi}
\end{center}
\begin{center}
Instituto de Astronomia,
Geof\'isica e Ci\^encias Atmosf\'ericas - \\University of S\~ao Paulo,
Brazil, Rua do Mat\~ao 1226, Cidade Universit\'aria, S\~ao Paulo,
05508-090
\end{center}
\begin{center}
{\bf Timothy C. Beers}
\end{center}
\begin{center}
CSCE: Center for the Study of Cosmic Evolution and
 JINA: \\Joint Institute for Nuclear Astrophysics, Michigan State
 University, East Lansing, MI 48824, USA
\end{center}

\begin{center}
 {\bf ABSTRACT}
\end{center}

 We present preliminary results for estimation of barium ([Ba/Fe]) and strontium ([Sr/Fe]) abundances ratios using
medium-resolution spectra (1-2 {\AA}). We established a calibration between the
abundance ratios and line indices for Ba and Sr, using
multiple regression and artificial neural network techniques. A comparison
between the two techniques (showing the advantage of the latter), as well as a
discussion of future work, is presented.
\section{Introduction}

Metal-poor stars (hereafter, MPS), defined as stars with less than 1/10th
of the solar iron abundance ( [Fe/H] $< -1.0$ ), are important for understanding
the chemical enrichment of the early Galaxy. However, the full range of
nucleosynthetic pathways for the production of the heaviest elements in such
stars is still not fully understood. Additional information
on abundance patterns are needed in order to provide clues and constraints on
the nature of element production for the first generations of stars.   

Many studies of MPS have been carried out in the last decades, making available
abundance patterns for stars with metallicities down to almost [Fe/H]$= -6$
(Frebel et al. 2008). However, a large fraction of these studies only provide
abundance values for elements lighter than iron. Among such studies, the
wide-angle spectroscopy surveys, such as the HK survey (Beers et al. 1985, 1992;
Beers et al. 1999), have performed an important role in the identification of
numerous low metallicity stars. The aim of the present project is to go one step
further and find a method to identify s-process-element enhanced MPS, using the
available medium-resolution spectroscopy data, and create a pre-selected sample
for detailed investigation at higher spectroscopic resolution.

Here we present preliminary results involving two methods for estimation of
barium ([Ba/Fe]) and strontium ([Sr/Fe]) abundances ratios, discuss the
difficulties encountered, and future developments.

\section{Data Samples}  
    
\begin{figure}[h]
\begin{center}
\includegraphics[width=7.65cm, angle=0]{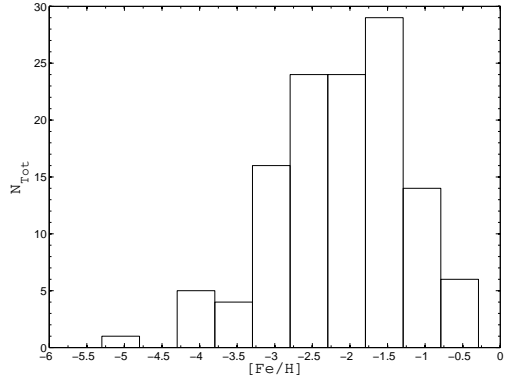}
\caption{\small{The metallicity distribution function of our calibration sample.}}\label{fehist}
\end{center}
\end{figure}

Our calibration data were selected from the library of medium
resolution spectra obtained during the HK (Beers et
al. 1985, 1992; Beers et al. 1999) and Hamburg/ESO (Christlieb 2003)
surveys with [Fe/H]$< -1.0$, and with Ba and Sr abundances ratios already
estimated by previous high-resolution studies. Their metallicity distribution
(derived from the high-resolution spectra) is shown in Figure (\ref{fehist}). We
have about 100 spectra with resolution between 1-2 {\AA}. 

We performed an extensive search for reported abundance ratio values
for [Ba/Fe] and [Sr/Fe] in the available literature. Since we found
differences from author to author as high as $\sim 0.2$ dex, we
decided to adopt a simple set of selection criteria for the adopted
abundances ratios instead of averaging all the values found for a single object,
in hopes to decreasing the overall errors.

The following criteria were applied to the literature sample:
\begin{itemize}

\item More recent values - from the nineties (late nineties were
  preferred) up to the present;
\item LTE values;
\item Highest S/N and resolving power.

\end{itemize}

\begin{figure}[h]
\begin{center}
\includegraphics[width=7.65cm, angle=0]{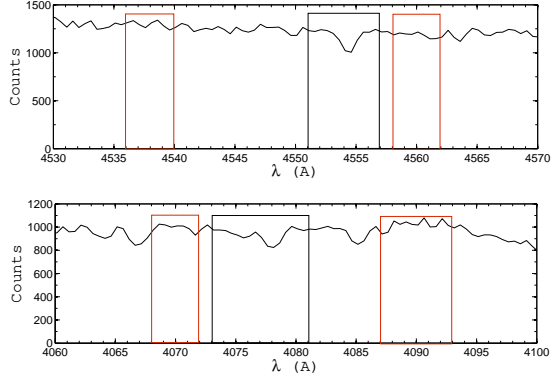}
\caption{\small{Definition of the indices Ba (top panel) and Sr (bottom panel) in the star LP 625-44. Red boxes
  are the sidebands and black box is the line band.}}\label{bain}
\end{center}
\end{figure}

Line indices for barium and strontium, as
defined in Table (\ref{indices}) and Figure (\ref{bain}), the Beers et
al. 1999 line index (KP) for the CaII K line, and the
2MASS\footnote[1]{Two Micron All Sky Survey} (Skrutskie et al. 2006)
near-IR colors $(J-K)_{0}$, were required for the application of our methods.
\begin{table}[h]
\begin{center}
\caption{Index definitions}\label{indices}
\begin{tabular}{|l||c|c|}
\hline Bands         & Ba        & Sr\\
\hline Blue Sideband & 4536-4540 & 4068-4072\\
       Line Band     & 4551-4557 & 4073-4081\\
       Red Sideband  & 4558-4562 & 4087-4093\\
\hline
\end{tabular}
\end{center}
\end{table}

Our program data contains about 35000 stars from the Sloan Digital Sky
Survey\footnote[2]{http://www.sdss.org/} Data Release 5 (SDSS; Gunn et
al. 1998; York et al. 2000; Adelman-McCarthy et al. 2007). These stars
were previously selected by metallicity, using the values of [Fe/H]
reported by the SDSS/SEGUE spectroscopic pipeline (Lee et al. 2008a,b;
Allende Prieto et al. 2008), have resolving power about $R=2000$, and
cover the wavelength range 3800 - 9200 {\AA}.

\section{Calibration}

\begin{figure}[h]
\begin{center}
\includegraphics[width=8.0cm, height=5.5cm,angle=0]{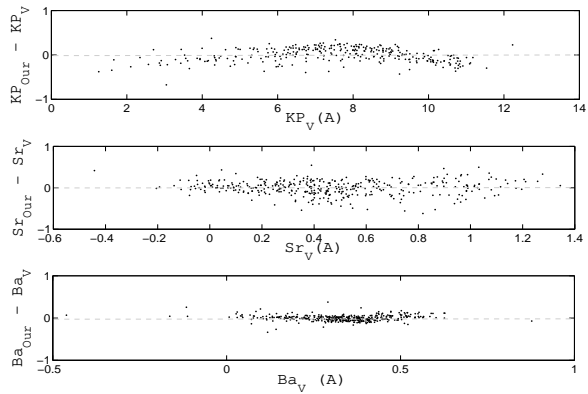}
\caption{\small{Comparison between the line indices measured with LECTOR
  and our code.}}\label{codigos}
\end{center}
\end{figure}

The line indices were measured using the LECTOR program by Vazdekis
(Vazdekis et al. 2003), as well as our own code. Figure (\ref{codigos}) shows
a comparison between the two different methods. The agreement between the
two codes is quite good, even for the Sr index which presents a larger
scatter. This larger scatter for the Sr index could be
  explained by two effects: a) The signal-to-noise ratio
  in this part of the spectrum -- According to Cayrel 1988 the
  uncertainty in the index is inversely proportional to the S/N ratio 
  thus, the smaller S/N in the Sr location increases the errors;
b) The continuum placement -- The Sr index band is closer to a strong
absorption line, which implies more difficulties in the
continuum determination. Therefore, the different procedures to estimate the
continuum by our and Vazdekis codes could also contribute to the
scatter. A quantitative analysis of the errors introduced by
continuum determination can be seen in Stetson \& Pancino 2008.   

Two different procedures were used to perform the calibration
-- multiple regression and application of an artificial neural network
(ANN), as described below.

\subsection{Multiple Regression}

We have obtained the abundance ratio [X/Fe], where X represents barium
or strontium, as a function of $log_{10}(X)$, $log_{10}(KP)$ and
$(J-K)_{0}$. The regression expressions for barium and strontium are
shown below:

\begin{eqnarray}
[Ba/Fe]&=& 7.34(0.02) + 4.37(0.02)log_{10}(Ba)\nonumber\\
       & & -3.71(0.03)log_{10}(KP)\nonumber\\
       & & -3.15(0.05)(J-K)_{o}.
\end{eqnarray}

\begin{eqnarray}
[Sr/Fe]&=& 4.59(0.05) + 1.96(0.05)log_{10}(Sr)\nonumber\\
       & & +0.04(0.09)log_{10}(KP)\nonumber\\
       & & -6.24(0.01)(J-K)_{o},
\end{eqnarray}
\noindent
where the values in parenthesis are the one sigma errors in the
determination of each coefficient.

\begin{figure}[h]
\begin{center}
\includegraphics[width=7.65cm, angle=0]{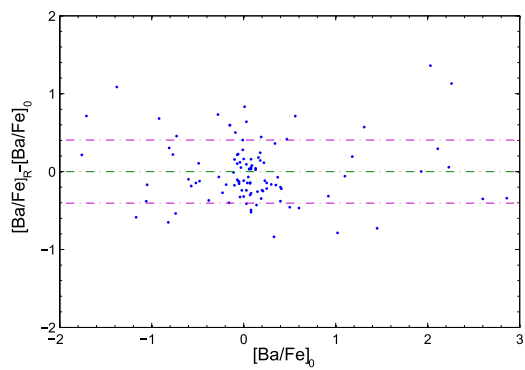}
\caption{\small{Distribution of the residual of the estimated [Ba/Fe]
    by regression,
    as compared to the adopted literature values. The green line
    represents the mean value of the residual and the pink lines are
    the 1$\sigma$ lines.}}\label{bareg}
\end{center}
\end{figure}

The standard deviation for the calibration values are about 0.4 dex and
1.1 dex for barium and strontium, respectively. Therefore, this method
is not reliable to measure the strontium abundance ratios, at least
by considering the present data. Figure (\ref{bareg})
shows the residual values for [Ba/Fe] as a function of the adopted literature values.

\section{Artificial Neural Networks}

Artificial neural networks (hereafter, ANNs) are playing an increasingly
important role in the analysis of large astronomical databases. This technique
can be used to find patterns, as well as make predictions (similar to
regression). In the latter case, the main advantage over simple regression is
the possibility for allowing non-linear interactions of the predictor variables,
which can change over the parameter space, in addition to the rapid training and
retraining.

For application of the ANN procedure, we have used the same input as in the
regression case, e. g., (Y, $X_{1}$, $X_{2}$, $X_{3}$) = ([X/Fe], $log_{10}(X)$,
$log_{10}(KP)$, $(J-K)_{0}$). Since we are conducting a supervised learning
exercise, we applied the back-propagation algorithm which propagates backwards
the error derivative of the weights, adjusting the weights in order to minimize
the final errors of determination. About 80 \% of the calibration sample were
used as a training set, setting aside 20 \% as a testing set, in a network with
2 layers and 10 nodes. In order to check the ANN's stability, we have used 10
different training and testing subsets. 

\begin{figure}[h]
\begin{center}
\includegraphics[width=7.65cm, angle=0]{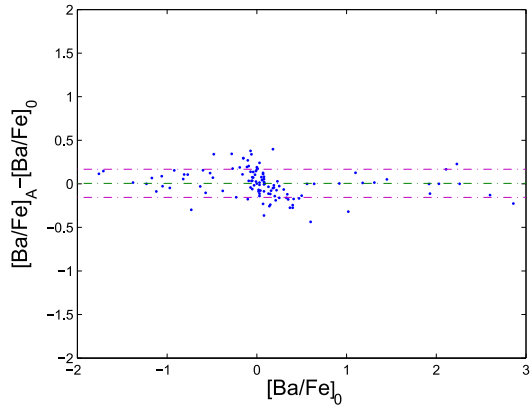}
\caption{\small{Distribution of the residual of the estimated [Ba/Fe]
    by ANN,
    as compared to the adopted literature values. The green line
    represents the mean value of the residual and the pink lines are
    the 1$\sigma$ lines.}}\label{baann}
\end{center}
\end{figure}

\begin{figure}[h]
\begin{center}
\includegraphics[width=7.65cm, angle=0]{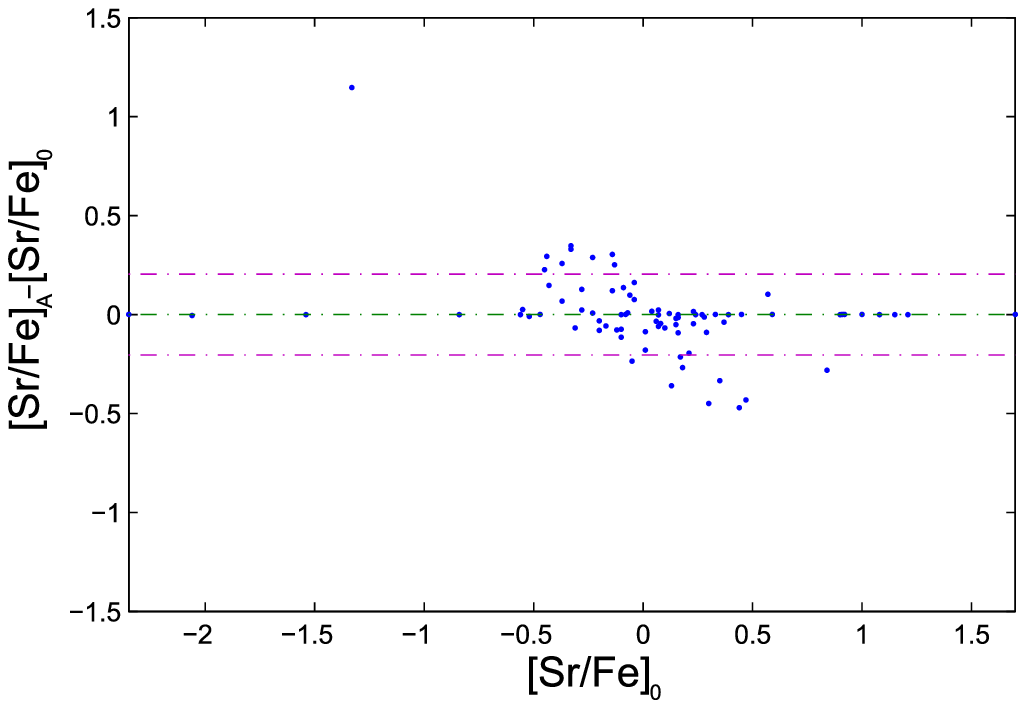}
\caption{\small{Distribution of the residual of the estimated [Sr/Fe],
    as compared to the adopted literature values. The green line
    represents the mean value of the residual and the pink lines are
    the 1$\sigma$ lines.}}\label{srann}
\end{center}
\end{figure}

\begin{figure}[htp]
\begin{center}
\includegraphics[width=7.65cm, angle=0]{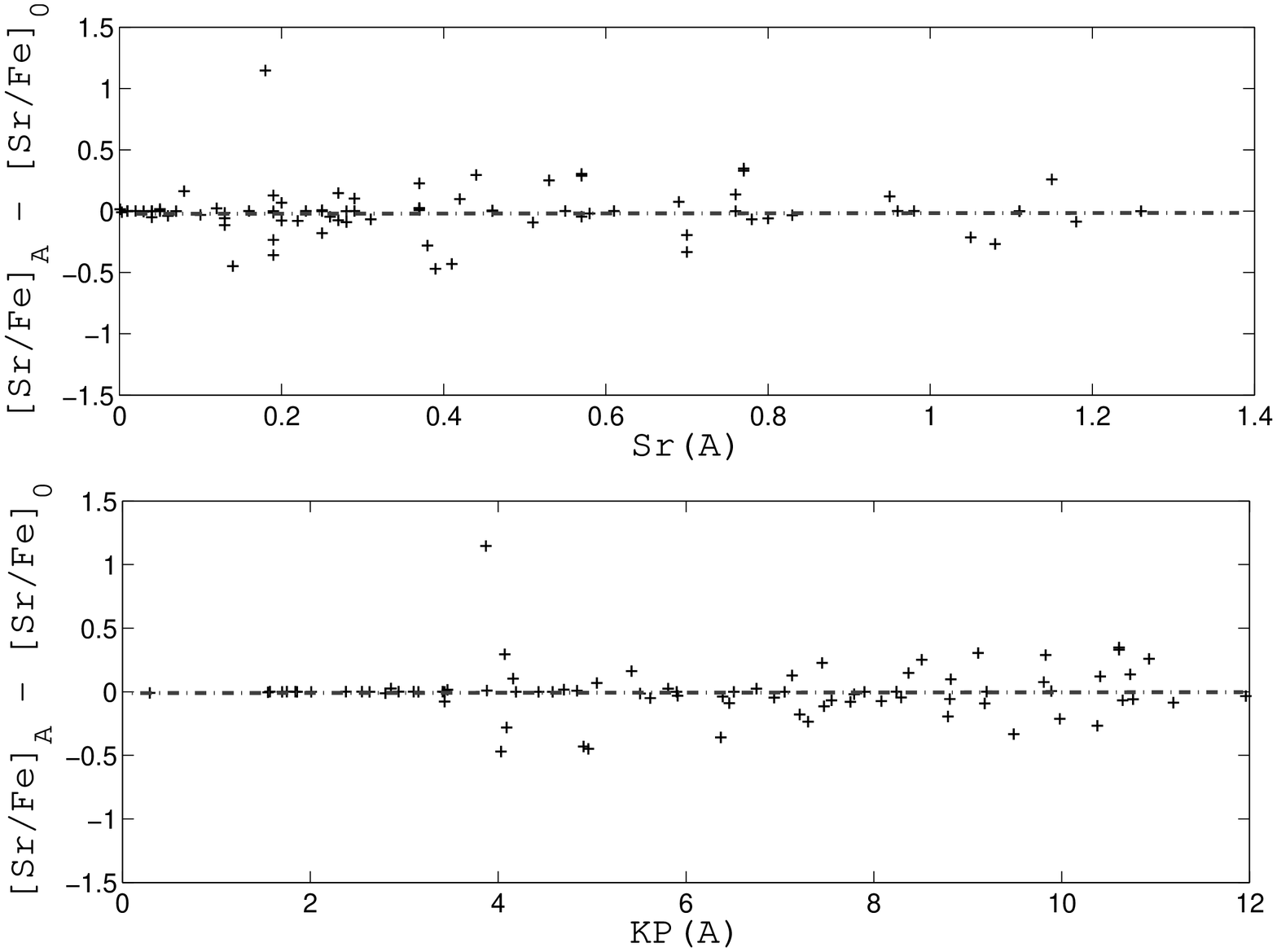}
\caption{\small{Distribution of the residual of the estimated [Sr/Fe]
    as a function of the predictor variables Sr (top panel) and KP
    (bottom panel).}}\label{srpred}
\end{center}
\end{figure}

\begin{figure}[htp]
\begin{center}
\includegraphics[width=7.65cm, angle=0]{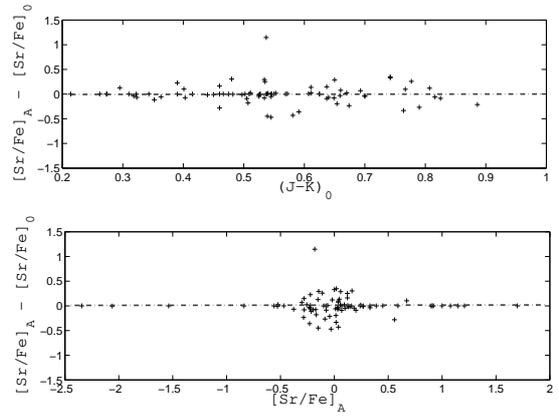}
\caption{\small{Distribution of the residual of the estimated [Sr/Fe]
    as a function of the predictor variable $(J-K)_{0}$ (top panel)
    and as function of the calculated values (bottom panel).}}\label{srpred2}
\end{center}
\end{figure}

Figures (\ref{baann}) and (\ref{srann}) show the results for the ANN
calibration. It is clear that considerable improvement has been
achieved, even for the Sr abundances, in the determination of
abundance ratios process. However, there are still
  problems to be solved in the Sr calibration. As can be seen in the
  residual plot, there is a small trend with 10\% - 20\%  error associated
  (not considering the outlier in the upper left). Since there is no
  trend when the residual is plotted versus the predictor variables
  (Sr, KP and $(J-K)_{0}$) and versus the calculated values (see
  Figures (\ref{srpred}) and (\ref{srpred2})), it is
  believed that a more homogeneous sample could help solve this problem.   

The standard deviation for these calibrations are 0.16 dex and 0.21 dex for Ba and
Sr, respectively.

\begin{figure}[htp]
\begin{center}
\includegraphics[width=7.65cm, height=4.5cm, angle=0]{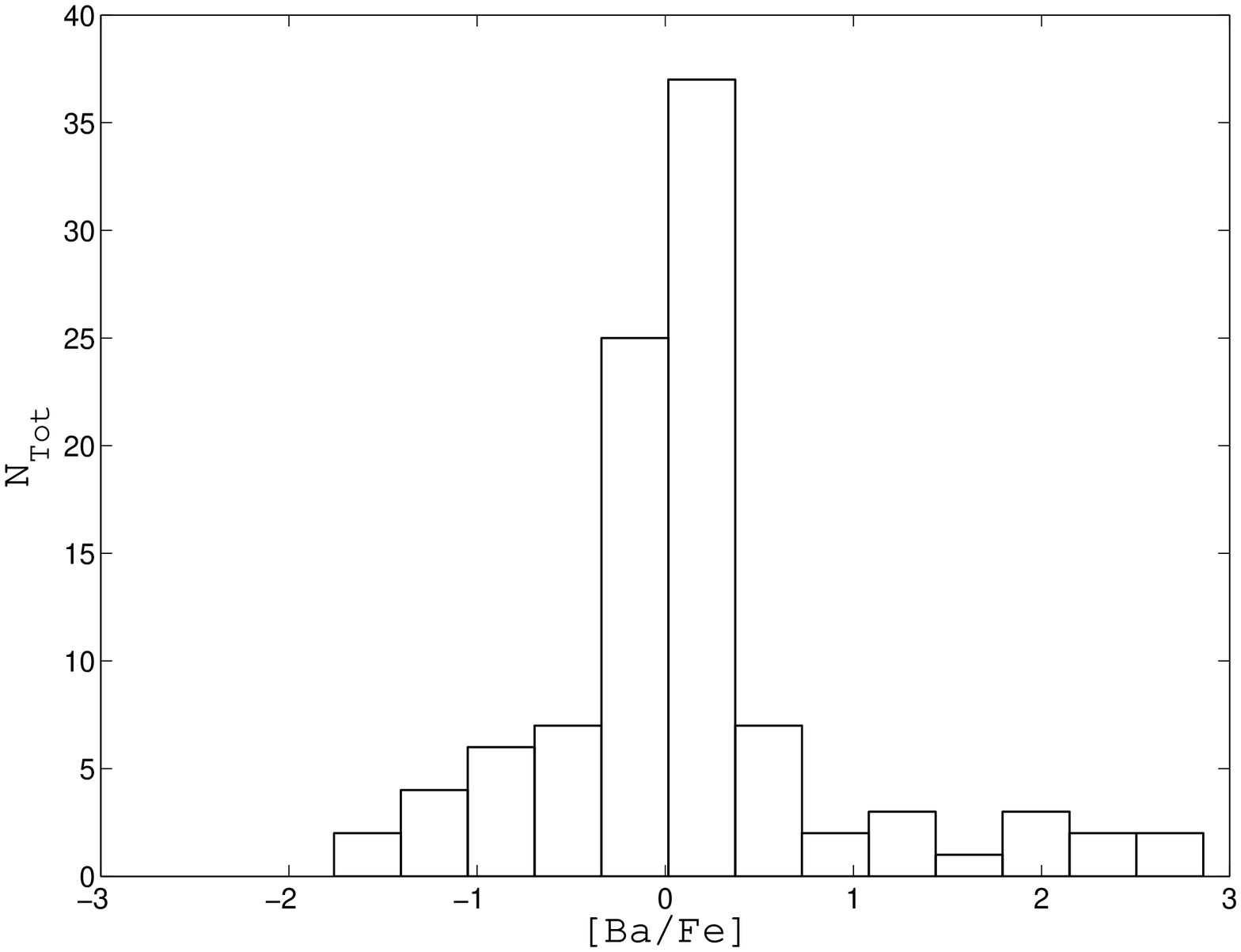}
\caption{\small{[Ba/Fe] distribution for the calibration sample.}}\label{bahist}
\end{center}
\end{figure}

\begin{figure}[htp]
\begin{center}
\includegraphics[width=7.65cm, height=4.5cm, angle=0]{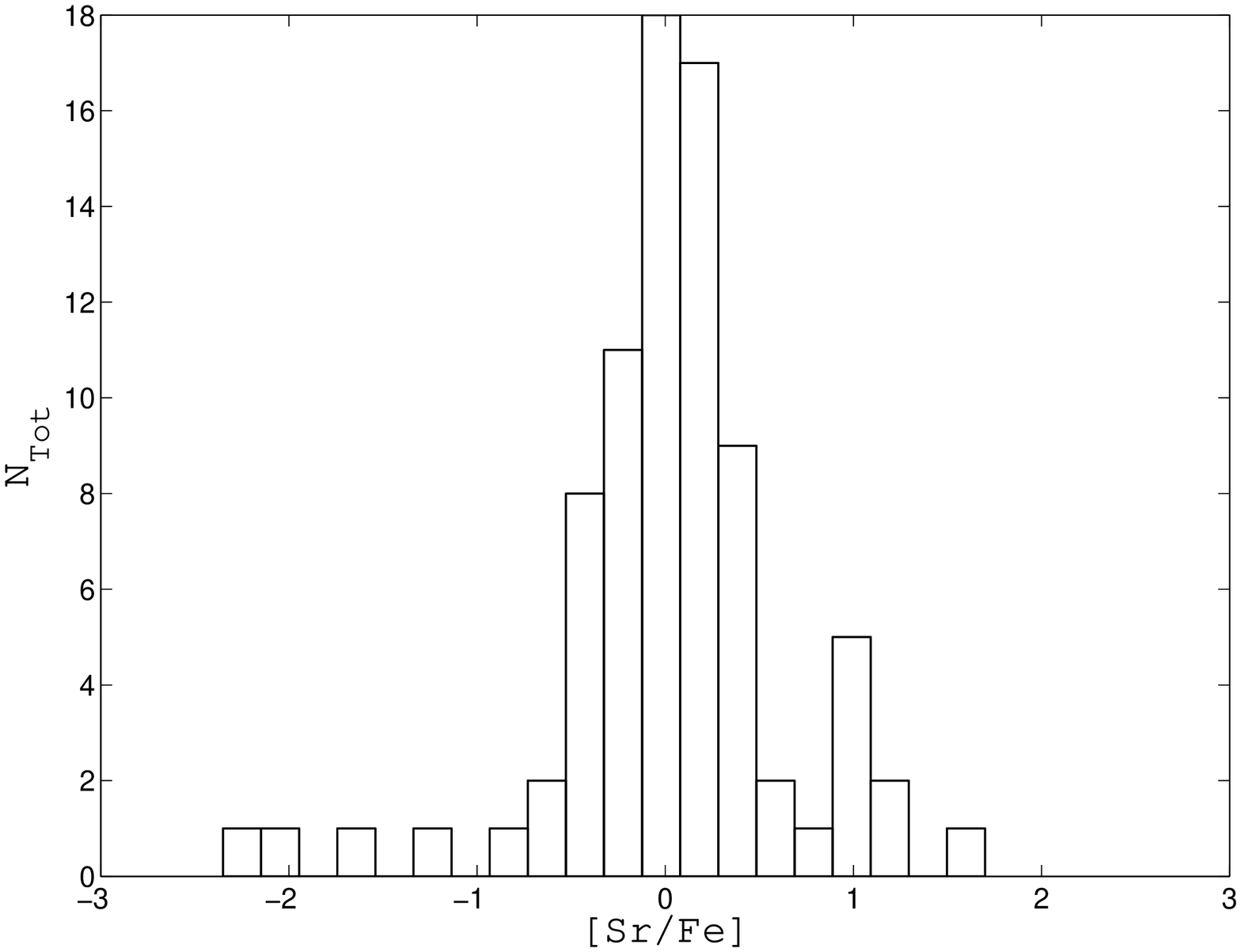}
\caption{\small{[Sr/Fe] distribution for the calibration sample.}}\label{srhist}
\end{center}
\end{figure}

The primary difficulty in performing such calibrations is related to the gaps in
the abundance space. As can be seen in Figures (\ref{bahist}) and
(\ref{srhist}), the distributions of stars, according to barium and strontium
abundance ratios, does not fill the entire parameter space. In particular, the
Sr abundances fall mostly in the range $-0.5 \le [Sr/Fe] \le 0.5$. This is
particularly important for the ANN approach, which is not suitable for
extrapolation.  

After considering these limitations, we have applied only the ANN method to the
program stars. The [Ba/Fe] and the [Sr/Fe] distributions for calibration stars
can be seen in Figures (\ref{bahist}) and (\ref{srhist}). It appears possible to
identify Ba-enhanced stars from medium-resolution data alone, with accuracy that
approaches that of high-resolution investigations. Our next goal is to increase
the calibration sample. There are about 100 additional stars whose Ba and Sr
abundances have already estimated using high-resolution spectroscopy data by
different groups, but which lack medium-resolution spectroscopy. The required
follow-up of the same objects with medium-resolution (R $\sim$ 2000)
spectroscopy is now underway. Application of the refined
estimation procedure to the program stars will appear in a forthcoming paper.

\section*{Acknowledgments} 
MAC would like to acknowledge the LOC and the Max-Planck-institut
f$\ddot{u}$r Astrophysik for making possible her attendance at the
workshop. MAC and SR thank the partial support from FAPESP, CNPq, and
Capes. TCB acknowledges support from the US National Science
Foundation under grants AST 04-06784 and AST 07-07776, as well as from
grants PHY 02-16783 and PHY 08-22648; Physics Frontier Center/Joint Institute for Nuclear Astrophysics (JINA).


\end{document}